# Theoretical analysis on x-ray cylindrical grating interferometer

Wenxiang Cong, Yan Xi, and Ge Wang

Department of Biomedical Engineering, Rensselaer Polytechnic Institute, Troy, NY 12180

**Abstract:** Grating interferometer is a state of art x-ray imaging approach, which can simultaneously acquire information of x-ray attenuation, phase shift, and small angle scattering. This approach is very sensitive to micro-structural variation and offers superior contrast resolution for biological soft tissues. The present grating interferometer often uses flat gratings, with serious limitations in the field of view and the flux of photons. The use of curved gratings allows perpendicular incidence of x-rays on the gratings, and gives higher visibility over a larger field of view than a conventional interferometer with flat gratings. In the study, we present a rigorous theoretical analysis of the self-imaging of curved transmission gratings based on Rayleigh-Sommerfeld diffraction. Numerical simulations have demonstrated the self-imaging phenomenon of cylindrical grating interferometer. The theoretical results are in agreement with the results of numerical simulations.

## 1. Introduction

X-ray phase-contrast imaging is to detect variations of the refractive index from distortions of a wavefront as it propagates through an object. The x-ray phase shift cross-section of low-Z elements is about a thousand times greater than those of attenuation within the diagnostic energy range, allowing a high signal-to-noise ratio and a high contrast resolution for biological tissue imaging [1]. X-ray phase-contrast imaging can be used to observe subtle structural variations of biological tissues, such as vessels, lesions, microcalcification, and fibrous tissues [2].

X-ray phase imaging methodology includes analyzer-based phase-contrast imaging, and free-space propagation methods, and grating interferometer imaging. Analyzer-based methods require high-precision crystals, a highly parallel and monochromatic x-ray beam [3]. Propagation-based methods need highly spatial coherent source, only available synchrotron sources or micro-focus x-ray sources with low power [4]. In contrast, grating-based phase imaging is the most promising with less of limitations. A coherent wavefront passing through a grating produces a Talbot self-image of the grating in the near field, allowing the reconstruction of the x-ray phase-contrast image. In 2003, a grating interferometer imaging system with a synchrotron x-ray source was setup in the hard x-ray region using a phase transmission grating and an analyzer absorption grating, and the feasibility of the x-ray grating phase imaging was demonstrated [5]. In 2006, based on the x-ray interferometer imaging, the synchrotron x-ray source was replaced by a coupling of a low-brilliance x-ray tube with a source grating to create an array of periodically repeating line sources, which individually emit sufficient spatial coherent x-ray beam but mutually incoherent for the x-ray phase-contrast imaging. This new design of the imaging system demonstrate that x-ray phase-contrast imaging can be efficiently performed with commonly x-ray source with an great potential for applications in biomedicine, industrial nondestructive testing, or security devices. The tomographic image reconstruction was also implemented with popular filtered back-projection algorithms to yield quantitative volumetric images of both the real and imaginary part of the refractive index of the sample [6, 7].

However, the field of view in the x-ray grating interferometer was in the order of some centimeters, which limit the application of x-ray grating phase imaging to small objects. Currently, the field of view in the x-ray grating interferometer was in the order of some centimeters, which limit the application of x-ray grating phase imaging to small objects. The present grating interferometer often uses flat gratings, which is always in favor of the imaging geometry with parallel beams. An x-ray tube is commonly a point source, which emits divergent beams in a cone shape. The flat grating imaging suffers from limitations in both the field of view and the flux of photons. The use of cylindrical gratings allows perpendicular incidence of x-rays on the gratings, and well matches the x-ray cone, leading to a higher visibility over a larger field of view than a conventional interferometer with flat gratings [8-10]. X-ray grating interferometer imaging is based on the Talbot effect, which is a self-imaging phenomenon under spatially coherent illumination. The theoretical analysis of Talbot effect for the plane periodic objects was studied assuming the Fresnel diffraction. Motivated by the absence of theoretical studies on curved periodic objects, in this paper, we present a rigorous formulation of the self-imaging with curved transmission gratings based on Rayleigh-Sommerfeld diffraction, and perform numerical simulation for the feasibility of the cylindrical grating interferometer imaging.

## 2. Self-imaging of cylindrical grating

A cylindrical phase grating consists of a large number uniformly spaced slits, which is periodic in term of the polar angle on the cylindrical surface with a radius of $r$. An x-ray tube is located at the center of the cylindrical grating, and assumed as a point source to emit a spherical wave, which can be described by [11]

$$S(\mathbf{r}) = \frac{1}{\|\mathbf{r}-\mathbf{r}_s\|} \exp(ik\|\mathbf{r}-\mathbf{r}_s\|) \quad (1)$$

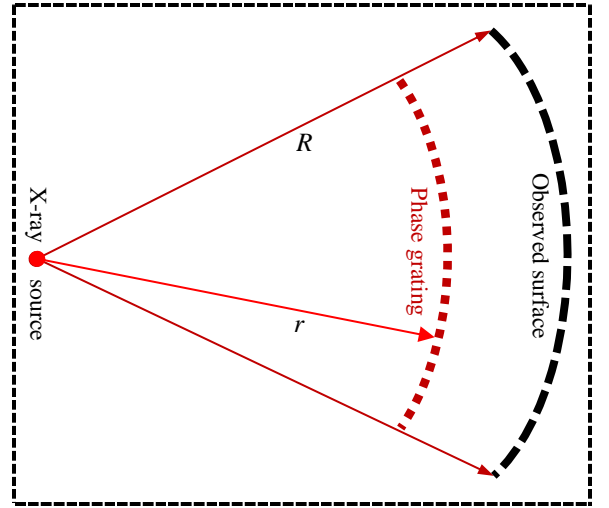

**Fig. 1.** *Curved grating imaging geometry*

where $k = \frac{2\pi}{\lambda}$ is the wave number for an x-ray wavelength $\lambda$, $\mathbf{r}$ is the observation position, and $\mathbf{r}_s = (x_s, y_s, 0)$ is the x-ray source position. With the setup of grating imaging system, as shown in Fig.1, the phase grating is curve with a radius $r$, and the incoming wave field at normal incidence to the curved phase grating, allowing to achieve a large field of view in a cone-beam geometry.

When the spherical wave propagates through a cylindrical grating with a period of $p$, the transmission function can be expressed by

$$T(\mathbf{r}_g) = S(\mathbf{r}_g) G(\mathbf{r}_g) \quad (2)$$

where $G(\mathbf{r}_g)$ is the transmission function of the phase grating. According to the Rayleigh-Sommerfeld diffraction integral, the observed wave field at a position $\mathbf{r}$ is described by [12]

$$Q(\mathbf{r}) = \frac{R-r}{i\lambda} \iint_\Sigma T(\mathbf{r}_g) \frac{\exp(ik\|\mathbf{r}-\mathbf{r}_g\|)}{\|\mathbf{r}-\mathbf{r}_g\|^2} d\Sigma \quad (3)$$

where $\Sigma$ is the curve surface of the cylindrical grating, $\mathbf{r}_g = (x_g, y_g, z_g)$ locates on the surface of the cylindrical grating, and $d = \|\mathbf{r}-\mathbf{r}_g\|$ is the distance between the point $\mathbf{r}_g$ and the observed position $\mathbf{r} = (x, y, z)$ located at another cylindrical surface with a radius of $R$ centered at the source position, paralleling the cylindrical phase grating. Using a variable change from Cartesian coordinates to cylindrical coordinates, $x_g - x_s = r\sin(\phi)$, $z_g = r\cos(\phi)$, $x - x_s = R\sin(\theta)$, $z = R\cos(\theta)$, $d = \sqrt{R^2 + r^2 - 2Rr\cos(\theta - \phi) + (y - y_g)^2}$, $-\frac{\pi}{2} < \phi < \frac{\pi}{2}$ and $-\frac{\pi}{2} < \theta < \frac{\pi}{2}$, Eq. (3) can be represented using cylindrical coordinate system,

$$Q(\theta, R, y) = \frac{R - r}{i\lambda} \exp(ikr) \int_{-\infty}^{\infty} \int_{-\frac{\pi}{2}}^{\frac{\pi}{2}} G(\phi, r, y_g) \frac{\exp(ikd)}{d^2} d\phi dy_g. \tag{4}$$

Furthermore, the periodic transmission function of the cylindrical phase grating can be expressed by Fourier series with a period of $p$ in polar coordinate system,

$$G(\phi, r, y_g) = \sum_{n=-\infty}^{\infty} c_n \exp\left(\frac{i2\pi n\phi}{p}\right) \tag{5}$$

Substituting Eq. (5) into Eq. (4), we have,

$$Q(\theta, R, y) = \frac{R - r}{i\lambda} \exp(ikr) \int_{-\infty}^{\infty} \int_{-\frac{\pi}{2}}^{\frac{\pi}{2}} \sum_{n=-\infty}^{\infty} c_n \exp\left(\frac{2\pi i n\phi}{p}\right) \frac{\exp(ikd)}{d^2} d\phi dy_g. \tag{6}$$

After changing the integration variable $\varphi = \phi - \theta$, we obtain,

$$Q(\theta, R, y) = \frac{(R - r)}{i\lambda} \exp(ikr) \sum_{n=-\infty}^{\infty} c_n \exp\left(\frac{2\pi i n\theta}{p}\right) \int_{-\infty}^{\infty} \int_{-\left(\frac{\pi}{2} + \theta\right)}^{\left(\frac{\pi}{2} - \theta\right)} \exp\left(\frac{2\pi i n\varphi}{p}\right) \frac{\exp(ikd)}{d^2} d\varphi dy_g, \tag{7}$$

with $d = \sqrt{R^2 + r^2 - 2Rr\cos(\varphi) + (y - y_g)^2}$. Clearly, Eq. (7) can be rewritten as,

$$\begin{cases} Q(\theta, R, y) = \sum_{n=-\infty}^{\infty} c_n b_n \exp\left(\frac{2\pi i n\theta}{p}\right) \\ b_n = \frac{p(R - r)}{i\lambda} \int_{-\infty}^{\infty} \int_{-\left(\frac{\pi}{2} - \theta\right)\frac{1}{p}}^{\left(\frac{\pi}{2} - \theta\right)\frac{1}{p}} \exp(2\pi i n\varphi) \frac{\exp(ikd)}{d^2} d\varphi dy_g \\ d = \sqrt{R^2 + r^2 - 2Rr\cos(p\varphi) + (y - y_g)^2} \end{cases} \tag{8}$$

From Eq. (8), if $b_n$ is a constant independent of $n$, the image on the observe surface is a period function same to the cylindrical phase grating. With the paraxial approximation $\cos(p\varphi) = 1 - \frac{1}{2}(p\varphi)^2$, we have

$$b_n = \frac{R - r}{i\lambda pRr} \int_{-\infty}^{\infty} \int_{-\left(\frac{\pi}{2} - \theta\right)\frac{1}{p}}^{\left(\frac{\pi}{2} - \theta\right)\frac{1}{p}} \exp(2\pi i n\varphi) \frac{\exp\left(i2\pi \tilde{k}\sqrt{\tilde{z}^2 + \varphi^2 + \tilde{y}^2}\right)}{\tilde{z}^2 + \varphi^2 + \tilde{y}^2} d\varphi d\tilde{y} \tag{9}$$

where, $\tilde{k} = \frac{p\sqrt{Rr}}{\lambda}$ $\tilde{y} = \frac{y - y_g}{p\sqrt{Rr}}$, and $\tilde{z} = \frac{R-r}{p\sqrt{Rr}}$. Because the grating period $p$ is very small, Eq. (9) can be considered as a 2D Fourier transform with respect to variable $\varphi$ and $\tilde{y}$. According to Wely's formula [11, 13],

$$-\frac{1}{2\pi}\iint \exp(i2\pi(\alpha\varphi + \beta\tilde{y}))\frac{\partial}{\partial \tilde{z}}\left(\frac{\exp\left(i2\pi\tilde{k}\sqrt{\tilde{z}^2 + \varphi^2 + \tilde{y}^2}\right)}{\sqrt{\tilde{z}^2 + \varphi^2 + \tilde{y}^2}}\right)d\varphi d\tilde{y} = \exp\left(i2\pi\tilde{z}\sqrt{\tilde{k}^2 - \alpha^2 - \beta^2}\right), \quad (10)$$

Eq. (9) has a closed-form,

$$b_n = \frac{1}{pRr}\exp\left(i2\pi\tilde{z}\sqrt{\tilde{k}^2 - n^2}\right) = \frac{1}{pRr}\exp\left(i2\pi\frac{R-r}{p\sqrt{Rr}}\sqrt{\left(\frac{p\sqrt{Rr}}{\lambda}\right)^2 - n^2}\right) \quad (11)$$

Furthermore, applying a binomial expansion to the square root in Eq. (11), we obtain,

$$b_n = \frac{1}{pRr}\exp\left(i2\pi\frac{R-r}{\lambda}\right)\exp\left(i\pi n^2\left(\frac{1}{r} - \frac{1}{R}\right)\frac{\lambda}{p^2}\right) \quad (12)$$

From Eq. (12), a characteristic distance can be defined to meet following equation,

$$\left(\frac{1}{r} - \frac{1}{R}\right) = \frac{mp^2}{\lambda}, \quad (13)$$

where $m$ is called as the fractional Talbot order. Thus, the wave field on the observed cylindrical surface with a radius of $R$ appears Talbot self-imaging at specific distance determined from Eq. (12) or Eq. (13) for the curve grating.

## 3. Numerical simulation

A point x-ray source with energy at 60keV emits spherical wave to propagate toward a cylindrical phase grating. The radius of cylindrical grating is equal to the distance to the source, which is set to 0.8m. The phase grating has a periodic of $4\times 10^{-6}$m and a duty cycle of 0.5 (line width to period ratio of the grating). In practice, commonly periodic structures are binary amplitude and phase gratings, which transmission function is described by

$$T(x) = \begin{cases} 1 & \text{if } 0 \leq (x \bmod p) < D \cdot p \\ Ae^{i\phi} & \text{if } D \cdot p \leq (x \bmod p) \leq p \end{cases}, \quad (14)$$

where mod is the modulo operator, $p$ is the periodic of grating, $D$ is the duty cycle defined as the ratio between the width of the grating lines and the period of the grating, and $\varphi$ is the phase shift. The image of the binary cylindrical grating with $A = 0$ is shown in Fig. 2(a), and the image at observed cylindrical surface with a radius of 0.95m is shown in Fig. 2(b), which Talbot self-imaging appears at the specific distance.

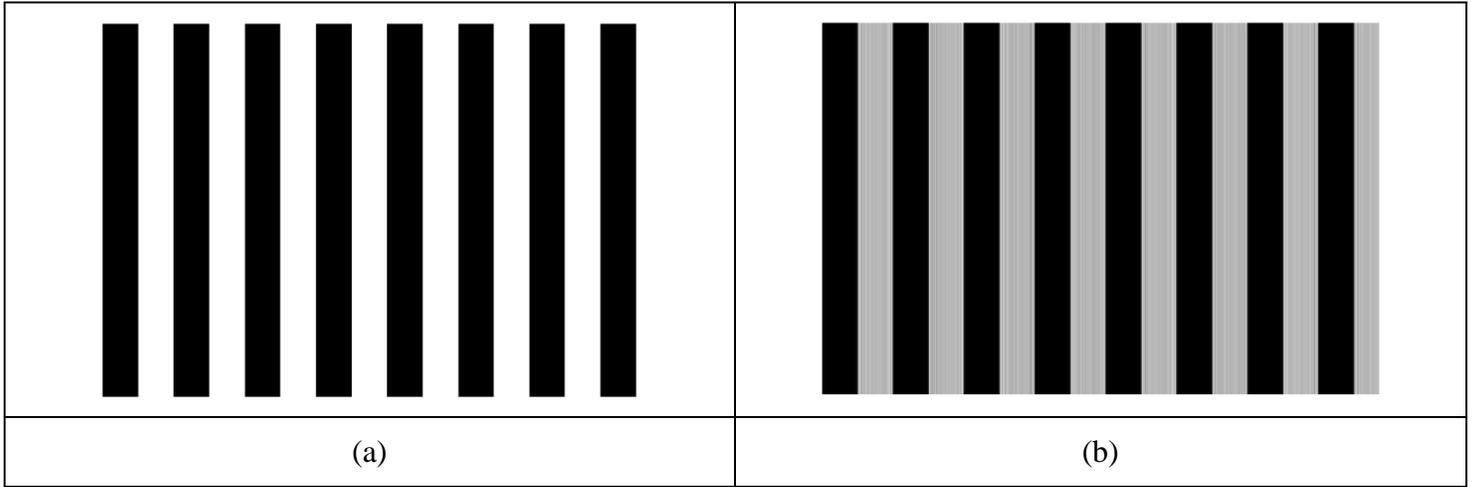

*Fig.2. Self-imaging simulation. (a) The image of the binary cylindrical grating; and (b) the image at observed cylindrical surface with a radius of 0.58m.*

The wave field downstream of a $\varphi = \pi/2$ phase shifting gratings and a $\phi = \pi$ phase shifting gratings are computed based on diffraction formula (8), and images of the wave field downstream are shown in Fig.3 (a-b). At a distance of the Talbot, the intensity distribution at specific distance is exactly same as transmission image of the grating.

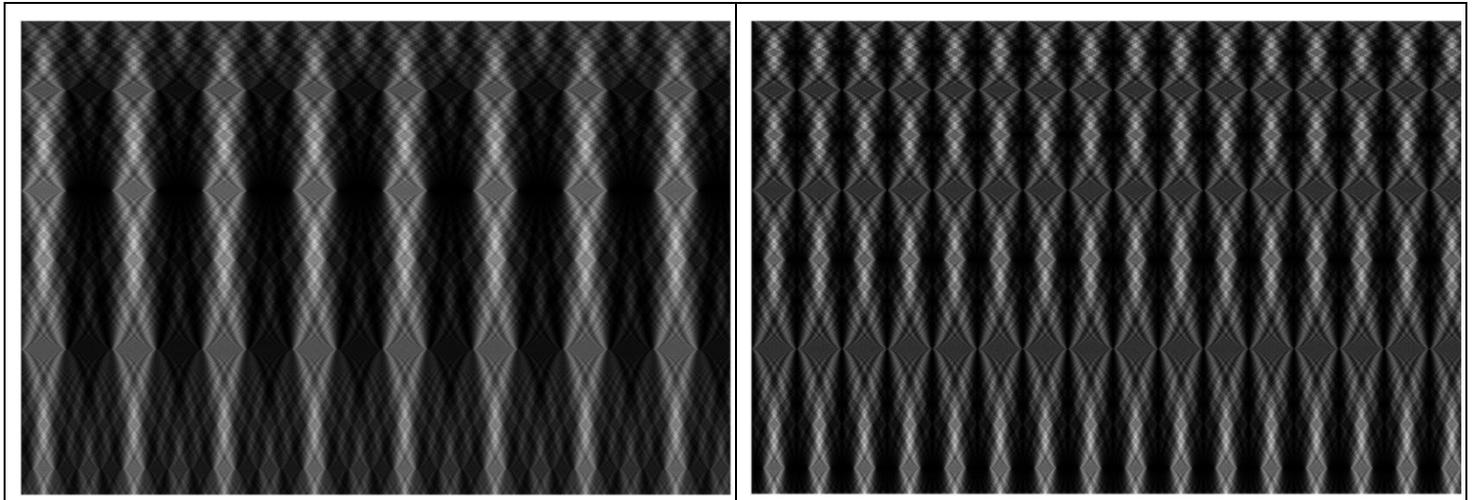

*Fig.3. The wave field downstream of a $\varphi = \pi/2$ phase shifting gratings and a $\phi = \pi$ phase shifting gratings.*

## 4. Discussions and Summary

X-ray grating interferometry can simultaneously acquire information of attenuation, refraction and scattering of the radiation, and is a promising imaging technique for clinical applications and industrial examination of samples. Impressive results have been reported for the grating-based phase imaging at low-energy x-ray and small field-of-view (FOV). For x-ray grating interferometry, high-energy x-ray and large FOV imaging are technically challenging. Because the common x-ray tube is a point source, the x-ray beam divergence will seriously limit the flux of photons, decrease visibility, and degrade the imaging performance in the case of flat gratings. With the use of cylindrical gratings, the x-ray beam enjoys perpendicular incidence on the cylindrical gratings, offering an opportunity to achieve a large field of view in a cone-beam geometry with a moderate tilting angle. The rigorous theoretical treatment has been reported for Talbot self-imaging with cylindrical gratings. It has been numerically shown that the wave field on the observed cylindrical surface does have the Talbot self-imaging patterns at the theoretically predicted specific distances. Similarly, the theoretical studies of Talbot self-imaging can be easily extended from 2D cylindrical gratings to 2D spherical gratings with a approximated periodic transmission. Relevant investigation is under way for subsequent publications

**Acknowledgments**

This work was supported by the National Institutes of Health (R01 EB016977).